# Atmospheric doping effects in epitaxial graphene: correlation of local and global electrical measurements

**Vishal Panchal[1], Cristina E. Giusca[1], Arseniy Lartsev[2], Nicholas A. Martin[1], Nathan Cassidy[1,3], Rachael L. Myers-Ward[4], D. Kurt Gaskill[4], and Olga Kazakova[1]***

[1]National Physical Laboratory, Teddington, TW11 0LW, UK
[2]Chalmers University of Technology, Gothenburg, S-412 96, Sweden
[3]University of Surrey, Guildford, Surrey, GU2 7XH, UK
[4]U.S. Naval Research Laboratory, Washington, DC 20375, USA

*E-mail: olga.kazakova@npl.co.uk

**Abstract.** We directly correlate the local (20-nm scale) and global electronic properties of a device containing mono-, bi- and tri-layer epitaxial graphene (EG) domains on 6*H*-SiC(0001) by simultaneously performing local surface potential measurements using Kelvin probe force microscopy and global transport measurements. Using well-controlled environmental conditions, where the starting state of the surface can be reproducibly defined, we investigate the doping effects of $N_2$, $O_2$, water vapour and $NO_2$ at concentrations representative of the ambient air. We show that presence of $O_2$, water vapour and $NO_2$ leads to p-doping of all EG domains. However, the thicker layers of EG are significantly less affected by the atmospheric dopants. Furthermore, we demonstrate that the general consensus of $O_2$ and water vapour present in ambient air providing majority of the p-doping to graphene is a common misconception. We experimentally show that even the combined effect of $O_2$, water vapour, and $NO_2$ at concentrations higher than typically present in the atmosphere does not fully replicate the state of the EG surface in ambient air. All doping effects can be reproducibly reversed by vacuum annealing. Thus, for EG gas sensors it is essential to consider naturally occurring environmental effects and properly separate them from those coming from targeted species.

## 1. Introduction

Graphene has attracted a great level of interest both from the research community and industry, owing to its novel physical properties with vast potential in technological applications as, for example, an ultrasensitive sensor module as part of a traditional silicon integrated circuit [1]. Epitaxial growth via thermal decomposition of SiC is a promising method for the production of large area graphene on a semi-insulating substrate [2]. The electronic properties of graphene are dependent on the layer thickness and dominated by interactions with its surroundings (i.e., the substrate and environment) [3,4]. For epitaxial graphene (EG) on SiC(0001), the interfacial layer (IFL) residing between the graphene layer and SiC comprises both $sp^2$ and $sp^3$ bonded carbon atoms [5,6], which induces strong n-doping to graphene. On the other hand, the surface of EG is sensitive to a range of gases present in the ambient air [7–12], where the strength of doping is highly dependent on their electronic properties. Understanding the influence of environmental doping on the fundamental electronic properties of EG is crucial for functionality of nanodevices for gas sensing and many other applications [1,13–15]. For effective functionality of an EG gas sensor, the sensor response originating from ambient air must be separated from the one coming from the targeted species.

While the p- and n-doping behaviour of small gas molecules has been widely reported mainly by global transport measurements on an entire device [12,14,16,17], these studies typically have not taken into account the local differences (i.e., thickness inhomogeneity) in the level of doping. Here, we perform unique simultaneous studies of the local variations of the electronic properties as manifested in the change of the surface potential ($V_{SP}$) and global transport properties in controlled environmental conditions. The simultaneous application of functional scanning probe microscopy (SPM) and global transport measurements enables the connection between nanoscale electronic properties and global, which has proven to be challenging in the past due to difficulties in reproducing exactly the same surface state of the sample for a given environmental condition.

Previous studies have shown that the water vapour present in the atmosphere acts as an electron acceptor, leading to an increase in p-doping (see, e.g. Ref. [8] and references within). In this work, we only address EG on SiC (i.e., intrinsically n-doped), as the mechanism of water interaction with graphene depends on many factors such as the type of graphene, chemistry, thermodynamics, etc [18,19]. For instance, it has been demonstrated that EG on SiC is less hydrophobic than free-standing graphene (i.e., exfoliated on a grid) [20], which is attributed to the presence of $sp^2$ and $sp^3$ carbons on IFL/SiC interface. Furthermore, a typical nominally mono-layer epitaxial graphene (1LG) sample contains patches of bi- (2LG) and tri-layers (3LG), where the thicker layers are more hydrophobic [21]. However, the amount of surface p-doping attained by domains of different thickness is unique due to the differences in hydrophobicity of 1-3LG.

Among other small molecules, $O_2$ and $NO_2$ has attracted substantial attention as potential p-dopant present in the ambient air [7,8,12,16,22]. For example, $O_2$ accounts for ~21% of ambient air by volume [23] and in cities with a heavy traffic, e.g. having a large number of diesel powered vehicles, the $NO_2$ levels can be within the range of 65-375 ppb [24]. For graphene devices, sensitivity to $O_2$ ~1.25% by volume and $NO_2$ down to sub-parts per billion (ppb) has been demonstrated [8,12,16,25].

In this paper, we study how the electronic properties of a 1-3LG dual cross Hall bar device fabricated out of epitaxial graphene on $6H$-SiC(0001) can be affected by various environmental conditions, namely: vacuum annealing, $N_2$, synthetic air (20% $O_2$ balance with $N_2$), water vapour (20-60% relative humidity (RH) balanced with synthetic air), $NO_2$ (60-480 parts per billion (ppb) balanced with synthetic air) and ambient air (Supporting Information Table S1). The specified concentrations of gases were selected to mimic their naturally occurring abundances and variations found in ambient air (Supporting Information Table S2) and the corresponding effects on EG. The Hall bar device was studied by mapping $V_{SP}$ with frequency-modulated Kelvin probe force microscopy (FM-KPFM) and global transport measurements by means of the AC Hall effect and channel resistance measurements.

## 2. Experimental Section
### 2.1. Graphene Synthesis
Semi-insulating $6H$-SiC(0001) commercial substrates (II-VI, Inc.) with resistivity >$10^{10}$ Ω cm were used in the work. The substrates were 8×8 mm$^2$ and misoriented ~0.05° from the basal plane, mainly in the (11-20) direction. Graphene was synthesized via Si sublimation from SiC using an overpressure of an inert gas. Prior to the growth, the substrate was etched in $H_2$ at 100 mbar using a ramp from room temperature to 1580 °C to remove polishing damage. At the end of the ramp, the $H_2$ was evacuated and Ar added to a pressure of 100 mbar (the transition takes about 2 minutes). Graphene was then synthesized at 1580 °C for 25 min in Ar. Afterwards, the sample was cooled in Ar to 800 °C [2]. Two or more layers of graphene formed using this procedure are proven to be Bernal stacked [26].

### 2.2. Device Fabrication
The Hall bar device was fabricated out of 1-3LG epitaxial graphene with electron beam lithography using PMMA and ZEP520 resists, oxygen plasma etching and electron beam physical vapour deposition of Cr/Au. For further details on the fabrication process, see Refs. [27] and [28]. The device fabrication process generally leads to contamination from resists and solvents. The residue layer has a thickness of 1-2 nm and has a tendency to blanket the graphene, which significantly affects the transport properties [27–29]. In order to investigate the purely environmental effects on the electronic properties of graphene, it is vital to remove the residues. This has been done by mechanically scraping the residues from side-to-side using contact-mode atomic force microscopy and soft cantilevers without damaging the graphene [28].

### 2.3. Global Transport and Surface Potential Measurements

Global transport measurements were accomplished using the AC Hall effect and 4-terminal resistance. Figure 1 shows the schematics of the AC Hall measurement setup [30]. See Supporting Information for details on the measurement.

The nanoscale electronic properties of epitaxial graphene were investigated using NT-MDT NTEGRA Aura SPM with Bruker PFQNE-AL probes. Figure 1 shows the single-pass frequency-modulated Kelvin probe force microscopy technique to map the surface electronic properties of a sample with spatial resolution comparable to the diameter of the probe apex (<20 nm) [31]. The $V_{SP}$ for 1-3LG was determined by taking the average value from representative areas of the Hall crosses and channel. The root mean square noise of the $V_{SP}$ measurements was determined in the range 15-45 mV; being largely dependent on the environmental conditions, e.g., mechanical properties of the probe, the gas flow rate. The thicknesses of the graphene layers were established by correlating Raman spectra of 1LG and 2LG to the $V_{SP}$. For details on the Raman spectra and maps performed on the same sample, see Ref. [21]. It should be noted that small 3LG domains were not observed on Raman maps due to low spatial resolution (~300 nm) of the system.

The presence of mixed number of layers is often unavoidable and rarely considered for gas sensing and electronic applications. In this situation, mapping the surface potential allows us to determine the graphene layer thickness and monitor the *local* electronic changes. For the device studied here, detailed analysis of the $V_{SP}$ map shows that cross 1 is covered by 70/30% of 1LG/2LG, cross 2 by 70/30% of 2LG/3LG and the 760 nm wide channel by 56/30/14% of 1LG/2LG/3LG (Figure 1 inset). For simplicity, from now on, Hall cross 1 and 2 will be referred to as *predominantly* 1LG and 2LG, respectively, and the channel as few-layer graphene (FLG). Thus, the carrier density of 1LG ($n_e^{1LG}$) is extracted from Hall measurements on cross 1 and similarly, the carrier density of 2LG ($n_e^{2LG}$) is from Hall measurements on cross 2 (see Section 2.1 in the Supporting Information for details on obtaining the carrier mobility). The transport measurements were performed continuously every minute and paused only for the duration of the FM-KPFM scans in order to avoid the device bias current affecting the surface potential of the graphene. Once the transport measurements had stabilized after each change to the environmental conditions, which usually occurred within 1-2 hours (unless stated otherwise), the FM-KPFM scans were performed. See Supporting Information for additional details on performing the measurement.

## 2.4. *Environmental Control*

The sample was mounted on a platinum thin film heater, placed on a ceramic TO-8 header. This allowed simultaneous measurement of the global transport properties and heating to 150 °C. The NT-MDT NTEGRA Aura SPM consisted of a vacuum tight chamber, which contained the TO-8 header assembly, allowing global transport measurements and simultaneous surface potential mapping of the 1-3LG Hall bar device in controlled environments. The SPM chamber was fitted with several inlet valves, allowing us to connect the turbomolecular pump, humidifier, electrical connections for global transport measurements and an outlet for venting the controlled environment inside the chamber. The turbomolecular and rotary pump combination evacuated the volume inside the chamber, allowing measurements in vacuum at pressures as low as $P = 1\times10^{-6}$ mbar. The mass flow controllers (MFCs), connected via the humidifier, enabled precise control of the gas concentration and humidity levels inside the chamber at standard temperature and pressure. The relative humidity inside the chamber was regulated by the humidifier which uses a heater to boil de-ionized water and mixes it with gas from the MFCs. The MFCs were connected to cylinders of $N_2$ (research grade at 99.9995% purity, where total impurities are 5 ppm), synthetic air (20% $O_2$ balanced with research grade $N_2$, where contamination from hydrocarbon is <0.1 ppm, $CO_2$ <1 ppm, $H_2O$ <2 ppm and $NO_X$ <0.1 ppm) and 120 ppm $NO_2$ balanced with synthetic air. All gas impurities are as quoted by the vendor, BOC Industrial Gases UK. The $N_2$ and synthetic air environment inside the SPM chamber was maintained by flowing in the respective gas at a rate of 2000 ml min$^{-1}$, whereas the environment of 60 ppb $NO_2$ balanced with synthetic air required 1 ml min$^{-1}$ flow of 120 ppm NO$_2$ and 2000 ml min$^{-1}$ flow of synthetic air. The measurements in ambient air were performed by flowing laboratory air through the chamber.

## 3. Results and Discussions

The paper is divided into three sections. In the first experiment, the sample conditioning was investigated to determine the pristine state of EG and reversibility of the EG surface properties upon exposure to ambient air. In the second experiment, we investigated the doping effects of majority constituents of ambient air (i.e., N$_2$, O$_2$ and water vapour) and in the third one – the effect of NO$_2$ at concentrations typically present in ambient air.

### 3.1. *Experiment 1: Sample conditioning and reversibility of EG surface*

The sample conditioning and reversibility of the EG surface state was tested by measuring the local $V_{SP}$, $n_e$, channel resistance ($R_{ch}$) and carrier mobility ($\mu_e$) of the FLG device during the ambient (laboratory) air–vacuum annealing–ambient air cycle. All measurements (if not indicated otherwise) were performed at 25 °C. The $V_{SP}$ maps presented in left panel of Figure 2a show that initially in ambient air (~40% RH), $V_{SP}$(1LG) < $V_{SP}$(2LG) ~ $V_{SP}$(3LG). Following on from $\Phi_{sample} = \Phi_{probe} − e\Delta V_{SP}$ [29,31], where $\Phi$ is the work function and $e$ is electron charge, $\Phi$ is lower for regions of higher $V_{SP}$. Therefore, in ambient air, the $\Phi$ is the largest for 1LG and the smallest for 2LG and 3LG.

The global transport properties presented in Figure 2b and 2c show that, initially in ambient air, the EG is n-type with the carrier density of 1LG and 2LG being $n_e^{1LG}$ = 1.19×10$^{12}$ cm$^{-2}$ and $n_e^{2LG}$ = 3.78×10$^{12}$ cm$^{-2}$, respectively. The resistance of the FLG channel was $R_{ch}$ = 9.5 kΩ, which, combined with the weighted arithmetic mean carrier density (see Supporting Information for additional details), translates to a channel weighted average carrier mobility of $\mu_e$ = 1440 cm$^2$V$^{-1}$s$^{-1}$. The sample was then annealed (150 °C) in vacuum ($P$ = 1×10$^{-6}$ mbar) for 7.4 hours and cooled back down to room temperature (25 °C) under vacuum, which resulted in a clear inversion of the $V_{SP}$ contrast, i.e., $V_{SP}$(1LG) > $V_{SP}$(2LG) > $V_{SP}$(3LG). Therefore, in this case, the $\Phi$ is the largest for 3LG and the smallest for 1LG (Figure 2a middle panel), i.e., opposite to what was observed in ambient air. The vacuum annealing process also dramatically increased the carrier density ($n_e^{1LG}$ = 1.10×10$^{13}$ and $n_e^{2LG}$ = 1.22×10$^{13}$ cm$^{-2}$) and decreased the $R_{ch}$ to 2.9 kΩ (at 150 °C) and further to 2.0 kΩ upon cooling to room temperature (25 °C). Simultaneously, the channel carrier mobility decreased to $\mu_e$ = 720 cm$^2$V$^{-1}$s$^{-1}$ at 150 °C and increased to $\mu_e$ = 1030 cm$^2$V$^{-1}$s$^{-1}$ upon cooling to room temperature.

The shift in Fermi energy is observed as a change in $\Phi$, which is directly related to $V_{SP}$. From the comparison of $V_{SP}$ and transport measurements, it should be noted that, although $\Phi$(1LG) < $\Phi$(2LG) after vacuum annealing, it does *not* necessarily imply $n_e^{1LG} > n_e^{2LG}$, as the band structures are fundamentally different for 1LG and 2LG, where the respective Fermi energy dispersion is [4]

$$E_F^{1LG} = v_F \hbar \sqrt{(\pi n_e)} \qquad (1)$$

and

$$E_F^{2LG} = \pi \hbar 2 n_e / 2 m_e^*, \qquad (2)$$

where $v_F$ - Fermi velocity, $\hbar$ - reduced Planck's constant and $m_e^*$ - effective mass of the electrons in 2LG. The effect of the band structure is clearly visible in Figure 2b, where the carrier density for 1LG is *always* less than 2LG.

Ambient air consists of a range of different gases present at various concentrations (Supporting Information Table S2) and these individual components can affect the electronic properties of graphene in their own unique way, and thus contribute to the overall doping of graphene from the exposed surface [8,16,22]. Annealing the sample in vacuum at 150 °C is a widely used process for desorbing molecular compounds from the surface of graphene [22,32–35], thus eliminating the doping effects of ambient air. The resulting increase in $n_e$ is a clear evidence of effective desorption of p-doping molecules (i.e., water vapour, $O_2$ and other gases) from the surface of graphene, thus, unmasking the inherent interactions between graphene and the underlying IFL [36,37]. Cooling the sample back down to room temperature, while maintaining the vacuum, has a negligible effect on $n_e$, which indicates that the level of internal doping is stable and no contamination occurs (i.e., from the chamber walls). However, the cooling effect further decreases $R_{ch}$ due to a weakening of the phonon scattering in epitaxial graphene via the LO phonon mode of 6$H$-SiC, thus directly affecting the carrier mobility in graphene (Figure 2c) [38]. These results clearly demonstrate that there are two competing parameters that govern the resistance and carrier mobility of graphene devices: carrier density and phonon scattering.

Exposing the sample again to a low flow of ambient air restored the $V_{SP}$ contrast, gradually decreased $n_e$, and increased $R_{ch}$ and $\mu_e$ over the course of 15 hours. However, the values initially observed in the ambient environment were not achieved within this limited period of time. These initial findings show that ambient air provides a significant amount of uncontrollable p-doping to EG, which can be effectively reversed by vacuum annealing. The latter procedure demonstrates a very good level of reversibility of molecular desorption and thus a good way to reproducibly clean the device.

### 3.2. Experiment 2: Effect of $N_2$, synthetic air and humidity

Here we investigate the device response to $N_2$, dry synthetic air and water vapour in synthetic air, which constitutes the largest components of ambient air. Prior to the start of the experiment, the sample was reverted to the pristine state by vacuum annealing and cooling it back down to room temperature under vacuum. The experiment was carried out by changing the environmental conditions in a stepwise manner: vacuum, dry $N_2$, dry synthetic air, 40-60-20% RH and ambient air. The specified order of relative humidity was elected to study any saturation effects. For each stepwise change in the environmental conditions, the $V_{SP}$ scans were performed once the transport properties had nearly reached saturation, which typically takes about 30 minutes, unless stated otherwise.

*3.2.1. Step 1: Vacuum annealing.* The vacuum annealing process was uses to return the sample to a well-defined pristine state (step 1 in Figure 3), where the $V_{SP}$ map and transport properties of the device were consistent with the previous cycle (i.e., $n_e^{1LG} = 1.09\times10^{13}$ cm$^{-2}$, $n_e^{2LG} = 1.26\times10^{13}$ cm$^{-2}$ and $\mu_e = 1006$ cm$^2$V$^{-1}$s$^{-1}$).

*3.2.2. Step 2: Pure $N_2$.* Rising the SPM chamber to atmospheric pressure by venting it to $N_2$ (step 2) affected the graphene such that the $V_{SP}$ contrast for $V_{SP}$(1LG) ~ $V_{SP}$(2LG) > $V_{SP}$(3LG) (Figure 3a and 3b), which is accompanied by a small decrease in the carrier density and mobility (Figure 3c). As $N_2$ is an inert gas and does not form a stable anion [39], it is unlikely to affect the electrical properties of graphene. However, the tubing that carries the gas from the cylinder to the SPM chamber probably holds water vapour and other contaminants, which are often difficult to remove with flushing alone. Therefore, this small amount of contamination is likely to cause minor p-doping observed with the introduction of $N_2$. This minor p-doping is sufficient to align the $\Phi$(1LG) and $\Phi$(2LG) in this environment (Figure 3a).

*3.2.3. Step 3: Synthetic air (20% $O_2$ balanced with $N_2$).* Further, we investigate the effect of $O_2$ on EG by mimicking the ambient air environment by introducing dry synthetic air. After introducing synthetic air, the $V_{SP}$ contrast difference between the graphene layers (1-2-3LG) was negligible (Figure 3a and 3b), which is the result of similar $\Phi$ for the graphene layers. The decrease in $n_e^{1LG}$ and $n_e^{2LG}$ (Figure 3c) is consistent with previous reports on the p-doping behaviour of $O_2$ [8,22], however, the effect is relatively minor. A further decrease in $\mu_e$ was again observed with the introduction of $O_2$. The fact that $n_e^{1LG}$ and $n_e^{2LG}$ are not equal at this stage, whereas the simultaneous $V_{SP}$ map reveals nearly identical work functions for 1-3LG, demonstrates the differences in the trends of $\Phi$ and $n_e$, which can be easily overlooked. With the additional knowledge of $n_e^{1LG}$ and $n_e^{2LG}$ from the simultaneous transport measurements and the fact that $E_F$ are aligned due to the electrical contact between 1-2LG, Equation 1 and Equation 2 were used to calculate the effective mass of the electrons in 2LG at step 3 ($n_e^{2LG} = 1.15\times10^{13}$ cm$^{-2}$) as $m_e^* = 0.0402\pm0.0005m_e$, where $m_e$ is the electron rest mass and the quoted uncertainty is 1σ standard deviation (see the Supporting Information for further details). It should be stressed that this value of $m_e^*$ is true only in this specific environmental condition and that it is similar to those quantified in previous experiments and theoretical calculations, where $m_e^*$ values in the range of 0.033-0.041$m_e$ were reported [40–44]. However, the experimental conditions in previous studies differ significantly from those in the present paper, e.g. either the carrier density was significantly lower in comparison to ours or the experiment was carried out at cryogenic temperatures.

*3.2.4. Steps 4, 5 and 6: Water vapour.* The effect of water vapour was further investigated in steps 4, 5 and 6, by changing the relative humidity inside the SPM chamber from 40% to 60% and then to 20%, respectively, while maintaining the background synthetic air environment (Figure 3a and 3b). Introducing 40% RH (i.e., typical ambient value) resulted in a clear reversal of the $V_{SP}$ contrast, i.e., back to the initial state where $V_{SP}$(1LG) < $V_{SP}$(2LG) ~ $V_{SP}$(3LG) (i.e., see Figure 2a). In addition, a further decrease in $n_e$ was observed (Figure 3c), where the decrease was more substantial for 1LG. In contrast to steps 2 and 3, introducing water vapour increased $\mu_e$, which may be a combination of reduced $n_e$ due to p-doping (yielding increased $\mu_e$) and the effect of water partly neutralizing some of the Coulomb scattering centres in the IFL, thereby reducing their scattering potential and increasing $\mu_e$ [45,46]. Increasing the relative humidity to 60%, followed by a decrease to 20%, made little change to the transport properties, which indicates that the initial level of water vapour (RH = 40 %) has quickly saturated (~1 hour) the graphene surface, forming water 'puddles' (see the right panel of Figure 2a, where the uneven contrast is attributed to formation of individual nanoscale 'puddles' of water on 3LG as opposed to the uniform contrast shown on 1LG indicating a lack of nanoscale water puddle formation); this would be consistent with the increased hydrophobicity of 3LG as compared to 1LG [29]. Whilst the $V_{SP}$ map reveals the formation of nanoscale 'puddles' of water, these localised features were *not* identifiable in the carrier density measurement alone due to averaging over the Hall cross, whereas the increase in carrier mobility suggests the presence of water on the surface of graphene [47]. Thus, water vapour significantly changes the properties of graphene, the effect saturates quickly and is strongly dependent upon the number of layers [10].

*3.2.5. Step 7: Re-exposing to ambient air.* Finally, exposing the device to ambient air (~40% RH) overnight (~15 hours) resulted in a dramatic decrease in carrier density ($n_e^{1LG}$ = 2.34×10$^{12}$ cm$^{-2}$ and $n_e^{2LG}$ = 8.02×10$^{12}$ cm$^{-2}$) and increase in mobility ($\mu_e$ = 1030 cm$^2$V$^{-1}$s$^{-1}$). However, despite the relatively long exposure to ambient air, the electronic properties still do not reach a stable state within this time scale, which is consistent with the findings of Ni *et al.* [10].

These measurements suggest that placing the graphene device in an environment containing $N_2$, $O_2$ and water vapour (either separately or in a mixture) will introduce p-doping and thus, decrease the overall electron carrier density of 1LG up to 47% as compared to the intrinsic value measured in vacuum (Table 1). Furthermore, the absolute change in $n_e$ is significantly less pronounced for 2LG, which shows that 1LG is ~2.6 times more susceptible to environmental changes (Table 1). However, despite the relatively high concentrations of $O_2$ and water vapour (mimicking ambient air), the further decrease in $n_e$ upon exposure to ambient air implies that there is additional p-doping from other gases, which are yet to be accounted for (Table S1).

### 3.3. Experiment 3: Effect of $NO_2$

We further investigated the effect of $NO_2$ on the same device by exposing the sample to 60-480 ppb $NO_2$, where 60 ppb is the highest recorded value since Jan 1, 2014 as measured by the Automatic Urban and Rural Network at the National Physical Laboratory site in Teddington, UK, where the experiment was performed. Prior to the measurements, the doping effect from the previous exposure was reversed by the vacuum annealing process described earlier. The electronic properties of the device in vacuum and synthetic air (Figure 4a, steps 1 and 2, respectively) were almost identical to the previous set of measurements, which again demonstrates the repeatability of the electronic properties. The introduction of 60 ppb dry $NO_2$ (step 3) and 60 ppb $NO_2$ with 40% RH (step 4) [both balanced with synthetic air] resulted in only a relatively minor decrease (increase) in $n_e$ ($\mu_e$), whereas exposing the sample again to ambient air (40% RH) and allowing it to stabilize overnight (~14 hours) shows once more a significant modification to the electronic properties (a decrease in $n_e^{1LG}$ from $6.66 \times 10^{12}$ cm$^{-2}$ to $1.62 \times 10^{12}$ cm$^{-2}$ for step 4 to 5). In a separate test, the effect of significantly larger concentrations of $NO_2$ (i.e., up to 480 ppb) in dry synthetic air was also investigated. Here, we observed an overall electron carrier density of 1LG decrease to $7.21 \times 10^{12}$ cm$^{-2}$ and $6.69 \times 10^{12}$ cm$^{-2}$ for 60 and 480 ppb $NO_2$ balanced with synthetic air, respectively, showing a relatively minor increase in the p-doping effect (Supporting Information Figure S1 and Table S3).

Throughout all the measurements presented in this work, we have empirically shown that, while $O_2$, water vapour and $NO_2$ provide p-doping to graphene, there is an additional 47% and 31.6% decrease in $n^{1LG}$ and $n^{2LG}$ (Figure 4a, step 4 to 5), respectively, that is yet to be accounted for, which is indicative of other p-doping molecules present in the ambient air that also interact with EG. Graphene and other carbon nanomaterials have also been shown to interact with p-dopants such as $N_2O_4$, $CO_2$ and hydrocarbons [17,25,48,49], which are likely candidates providing the additional p-doping when the EG device was exposed to ambient air. The physical origin of the molecular p-doping is based on electron charge transfer from graphene to the molecule (i.e., $H_2O$, $O_2$, $CO_2$, etc.), which occurs when the lowest unoccupied molecular orbital is below the Dirac point [7]. However, for water in particular, the exact mechanism is being debated as a complete understanding of the water-graphene interaction is still lacking.

It is well-known that small gas molecules, such as $H_2$, $O_2$ and $H_2O$, can be intercalated in between the IFL and SiC. However, the intercalation process typically requires temperatures around 500-1100 °C and 30-900 mbar gas pressure [50–52]. In the present work, the sample was only heated to 150 °C and in a *vacuum*. These conditions are highly unsuitable for the intercalation process, i.e., absence of intercalation gas and significantly lower temperatures. Thus, it is very unlikely that such intercalation could occur in the present experiment. Very high degree of reproducibility of the results also speaks against possible intercalation.

Our results show that the changes to the environmental conditions have resulted in a noticeably larger variation of the electronic properties for 1LG than 2LG/3LG, which suggests that 1LG is significantly more prone to atmospheric doping. A possible reason for this could be the preferential attachment of molecules to 1LG due to structural defects. However, scanning tunnelling microscopy and transmission electron microscopy studies on EG in literature [53–55] and the low intensity of the Raman D-peak on our sample have revealed a relatively defect free surface (Supporting Information Figure S2). Alternatively, electric field screening, which having a strong effect in *AB*-stacked 2LG on SiC [26,27], could explain the increased sensitivity of 1LG to doping. For example, the graphene-molecule interactions are at least partly governed by the charge state of the substrate, where in the present case the IFL provides strong n-doping to EG [5,6]. Figure 4b shows a schematic representation of the electric field between 1-2-3LG as a result of the n-doping provided by the IFL. In this system, the top layer in the 2LG stack can screen the charge associated with the IFL from the molecules more effectively than 1LG, which reduces the graphene-molecule electrostatic interaction and thus, reduces the effectiveness of molecular doping on 2LG (Figure 4b). The screening behaviour will be even more prominent for 3LG where higher hydrophobicity leads to the formation of water 'puddles'. While in our measurements we cannot directly confirm the role of structural defects, the effect of electrical screening is clearly observed and plays a significant role in the system described here.

**Conclusion**
We performed novel measurements of the surface potential, carrier density and carrier mobility to simultaneously investigate the local and global electronic properties of a 1-3LG device in controlled environments. In ambient air, the device exhibited a relatively low carrier density, high carrier mobility and work function that decreases with the layer thickness. However, annealing the sample in vacuum at 150 °C reversibly reproduces the cleanest state by eliminating atmospheric doping of EG with carrier density reaching a maximum value of $n_e^{1LG}$ = 1.10×10$^{13}$ and $n_e^{2LG}$ = 1.22×10$^{13}$ cm$^{-2}$. Furthermore, the distinct work function contrast was completely inverted in vacuum, which further indicates a significant modification of the transport properties of EG as compared to ambient air. We have experimentally shown that $O_2$, water vapour and $NO_2$ are only partially responsible for lowering of the electron carrier density of the EG as it is observed in ambient air. These results suggest that there are still moieties in the ambient air which result in significant amount of p-doping (additional 47.4% and 31.6% decrease in $n^{1LG}$ and $n^{2LG}$, respectively); potential sources are positively charged species such as $N_2O_4$, $CO_2$ and hydrocarbons, which may result in electron withdrawal. The atmospheric effects are significantly less pronounced for 2LG/3LG than for 1LG, as the top graphene layer(s) likely screens the charge from the atmospheric dopants. The unique combination of surface potential and global transport measurements also enabled us to determine the effective mass of the electron in 2LG as $m_e^*$ ~0.0402±0.0005$m_e$, which is otherwise not possible from carrier density or surface potential alone. Thus, we show that the electronic properties of EG are prone to change under combined action of different gases typically found in the ambient air. These species have a varying degree of doping effect on EG, which makes it difficult to distinguish the changes from any one type of molecule. Clearly, narrowing the sensitivity of graphene to specific target molecules is key to achieving optimum molecular sensing performance.

**Acknowledgements**
Authors acknowledge support of EC grants Graphene Flagship (No. CNECT-ICT-604391), EMRP under project GraphOhm (No 117359) and NMS under the IRD Graphene Project (No 118616). Work at the U.S. Naval Research Laboratory was supported by the Office of Naval Research. The authors are grateful to Ivan Rungger and David Carey for useful discussions.

**Supporting Information**
Supporting Information is available from the www.iopscience.iop.org or from the authors.

**References**
[1] Ferrari A C, Bonaccorso F, Falko V, Novoselov K S, Roche S, Bøggild P, Borini S, Koppens F, Palermo V, Pugno N, Garrido J A, Sordan R, Bianco A, Ballerini L, Prato M, Lidorikis E, Kivioja J, Marinelli C, Ryhänen T, Morpurgo A, Coleman J N, Nicolosi V, Colombo L, Fert A, Garcia-Hernandez M, Bachtold A, Schneider G F, Guinea F, Dekker C, Barbone M, Galiotis C, Grigorenko A, Konstantatos G, Kis A, Katsnelson M, Beenakker C W J, Vandersypen L, Loiseau A, Morandi V, Neumaier D, Treossi E, Pellegrini V, Polini M, Tredicucci A, Williams G M, Hong B H, Ahn J H, Kim J M, Zirath H, van Wees B J, van der Zant H, Occhipinti L, Di Matteo A, Kinloch I A, Seyller T, Quesnel E, Feng X, Teo K, Rupesinghe N, Hakonen P, Neil S R T, Tannock Q, Löfwander T and Kinaret J 2014 Science and technology roadmap for graphene, related two-dimensional crystals, and hybrid systems *Nanoscale* **7** 4598–810

[2] Nyakiti L O, Wheeler V D, Garces N Y, Myers-Ward R L, Eddy C R and Gaskill D K 2012 Enabling graphene-based technologies: Toward wafer-scale production of epitaxial graphene *MRS Bull.* **37** 1149–57

[3] Castro Neto A H, Peres N M R, Novoselov K S and Geim A K 2009 The electronic properties of graphene *Rev. Mod. Phys.* **81** 109–62

[4] Das Sarma S, Adam S, Hwang E and Rossi E 2011 Electronic transport in two-dimensional graphene *Rev. Mod. Phys.* **83** 407–70

[5] Kopylov S, Tzalenchuk A, Kubatkin S and Fal'ko V I 2010 Charge transfer between epitaxial graphene and silicon carbide *Appl. Phys. Lett.* **97** 112109

[6] Farmer D B, Perebeinos V, Lin Y-M, Dimitrakopoulos C and Avouris P 2011 Charge trapping and scattering in epitaxial graphene *Phys. Rev. B* **84** 205417

[7] Leenaerts O, Partoens B and Peeters F M 2008 Adsorption of H2O, NH3, CO, NO2, and NO on graphene: A first-principles study *Phys. Rev. B* **77** 125416

[8] Kong L, Enders A, Rahman T S and Dowben P A 2014 Molecular adsorption on graphene *J. Phys. Condens. Matter* **26** 443001

[9] Novikov S, Satrapinski A, Lebedeva N and Iisakka I 2013 Sensitivity Optimization of Epitaxial Graphene-Based Gas Sensors *IEEE Trans. Instrum. Meas.* **62** 1859–64

[10] Ni Z H, Wang H M, Luo Z Q, Wang Y Y, Yu T, Wu Y H and Shen Z X 2010 The effect of vacuum annealing on graphene *J. Raman Spectrosc.* **41** 479–83

[11] Yang Y, Brenner K and Murali R 2012 The influence of atmosphere on electrical transport in graphene *Carbon N. Y.* **50** 1727–33

[12] Novikov S, Lebedeva N and Satrapinski A 2015 Ultrasensitive NO 2 Gas Sensor Based on Epitaxial Graphene *J. Sensors* **2015** 108581

[13] Fowler J D, Allen M J, Tung V C, Yang Y, Kaner R B and Weiller B H 2009 Practical Chemical Sensors from Chemically Derived Graphene *ACS Nano* **3** 301–6

[14] Dan Y, Lu Y, Kybert N J, Luo Z and Johnson a T C 2009 Intrinsic response of graphene vapor sensors. *Nano Lett.* **9** 1472–5

[15] Schedin F, Geim A K, Morozov S V, Hill E W, Blake P, Katsnelson M I and Novoselov K S 2007 Detection of individual gas molecules adsorbed on graphene *Nat. Mater.* **6** 652–5

[16] Yuan W and Shi G 2013 Graphene-based gas sensors *J. Mater. Chem. A* **1** 10078

[17] Yoon H J, Jun D H, Yang J H, Zhou Z, Yang S S and Cheng M M C 2011 Carbon dioxide gas sensor using a graphene sheet *Sensors Actuators, B Chem.* **157** 310–3

[18] Henderson M 2002 The interaction of water with solid surfaces: fundamental aspects revisited *Surf. Sci. Rep.* **46** 1–308

[19] Cao P, Xu K, Varghese J O and Heath J R 2011 The microscopic structure of adsorbed water on hydrophobic surfaces under ambient conditions. *Nano Lett.* **11** 5581–6

[20] Zhou H, Ganesh P, Presser V, Wander M C F, Fenter P, Kent P R C, Jiang D, Chialvo A A, McDonough J, Shuford K L and Gogotsi Y 2012 Understanding controls on interfacial wetting at epitaxial graphene: Experiment and theory *Phys. Rev. B* **85** 035406

[21] Munz M, Giusca C E, Myers-Ward R L, Gaskill D K and Kazakova O 2015 Thickness-Dependent Hydrophobicity of Epitaxial Graphene *ACS Nano* **9** 8401–11

[22] Pinto H, Jones R, Goss J P and Briddon P R 2010 Mechanisms of doping graphene *Phys. Status Solidi* **207** 2131–6

[23] Lide D R, Data S R, Board E A, Baysinger G, Chemistry S, Library C E, Berger L I, Goldberg R N, Division B, Kehiaian H V, Kuchitsu K, Rosenblatt G, Roth D L and Zwillinger D 2005 *CRC Handbook of Chemistry and Physics, Internet Version 2005*

[24] Sweeney B P, Quincey P G, Green D and Fuller G W 2015 Quantifying the impact of nitric oxide calibration gas mixture oxidation on reported nitrogen dioxide concentrations *Atmos. Environ.* **105** 169–72

[25] Wehling T O, Novoselov K S, Morozov S V., Vdovin E E, Katsnelson M I, Geim A K and Lichtenstein A I 2008 Molecular doping of graphene *Nano Lett.* **8** 173–7

[26] Ohta T, Bostwick A, McChesney J, Seyller T, Horn K and Rotenberg E 2007 Interlayer Interaction and Electronic Screening in Multilayer Graphene Investigated with Angle-Resolved Photoemission Spectroscopy *Phys. Rev. Lett.* **98** 16–9

[27] Panchal V, Giusca C E, Lartsev A, Yakimova R and Kazakova O 2014 Local electric field screening in bi-layer graphene devices *Front. Phys.* **2**

[28] Panchal V 2014 *Epitaxial graphene nanodevices and their applications for electronic and magnetic sensing (Ph.D. Thesis)* (Royal Holloway, University of London)

[29] Kazakova O, Panchal V and Burnett T 2013 Epitaxial Graphene and Graphene–Based Devices Studied by Electrical Scanning Probe Microscopy *Crystals* **3** 191–233

[30] Tousson E and Ovadyahu Z 1988 Anomalous field dependence of the Hall coefficient in disordered metals *Phys. Rev. B* **38** 12290–7

[31] Panchal V, Pearce R, Yakimova R, Tzalenchuk A and Kazakova O 2013 Standardization of surface potential measurements of graphene domains *Sci. Rep.* **3** 2597

[32] Lin Y-C, Lu C-C, Yeh C-H, Jin C, Suenaga K and Chiu P-W 2012 Graphene annealing: how clean can it be? *Nano Lett.* **12** 414–9

[33] Sikora A, Woszczyna M, Friedemann M, Ahlers F J and Kalbac M 2012 AFM diagnostics of graphene-based quantum Hall devices *Micron* **43** 479–86

[34] Lohmann T, von Klitzing K and Smet J H 2009 Four-terminal magneto-transport in graphene p-n junctions created by spatially selective doping. *Nano Lett.* **9** 1973–9

[35] Chan J, Venugopal A, Pirkle A, McDonnell S, Hinojos D, Magnuson C W, Ruoff R S, Colombo L, Wallace R M and Vogel E M 2012 Reducing extrinsic performance-limiting factors in graphene grown by chemical vapor deposition. *ACS Nano* **6** 3224–9

[36] Emtsev K V., Seyller T, Speck F, Ley L, Stojanov P, Riley J D and Leckey R C G 2007 Initial Stages of the Graphite-SiC(0001) Interface Formation Studied by Photoelectron Spectroscopy *Mater. Sci. Forum* **556-557** 525–8

[37] Emtsev K V., Speck F, Seyller T and Ley L 2008 Interaction, growth, and ordering of epitaxial graphene on SiC{0001} surfaces: A comparative photoelectron spectroscopy study *Phys. Rev. B* **77** 155303

[38] Lara-Avila S, Tzalenchuk A, Kubatkin S, Yakimova R, Janssen T J B M, Cedergren K, Bergsten T and Fal'ko V 2011 Disordered Fermi Liquid in Epitaxial Graphene from Quantum Transport Measurements *Phys. Rev. Lett.* **107** 166602

[39] Rienstra-Kiracofe J C, Tschumper G S, Schaefer H F, Nandi S and Ellison G B 2002 Atomic and Molecular Electron Affinities: Photoelectron Experiments and Theoretical Computations *Chem. Rev.* **102** 231–82

[40] García-Flores A F, Terashita H, Granado E and Kopelevich Y 2009 Landau levels in bulk graphite by Raman spectroscopy *Phys. Rev. B* **79** 113105

[41] Zou K, Hong X and Zhu J 2011 Effective mass of electrons and holes in bilayer graphene: Electron-hole asymmetry and electron-electron interaction *Phys. Rev. B - Condens. Matter Mater. Phys.* **84** 085408

[42] Rutter G M, Jung S, Klimov N N, Newell D B, Zhitenev N B and Stroscio J A 2011 Microscopic Polarization in Bilayer Graphene *Nat. Phys.* **7** 649–55

[43] Huang J, Alexander-Webber J a, Janssen T J B M, Tzalenchuk A, Yager T, Lara-Avila S, Kubatkin S, Myers-Ward R L, Wheeler V D, Gaskill D K and Nicholas R J 2015 Hot carrier relaxation of Dirac fermions in bilayer epitaxial graphene *J. Phys. Condens. Matter* **27** 164202

[44] Koshino M and Ando T 2006 Transport in bilayer graphene: Calculations within a self-consistent Born approximation *Phys. Rev. B* **73** 245403

[45] Lu J, Pan J, Yeh S-S, Zhang H, Zheng Y, Chen Q, Wang Z, Zhang B, Lin J-J and Sheng P 2014 Negative correlation between charge carrier density and mobility fluctuations in graphene *Phys. Rev. B* **90** 085434

[46] Murat A, Rungger I, Sanvito S and Schwingenschlögl U 2015 Mechanism of $H_2O$-Induced Conductance Changes in $AuCl_4$-Functionalized CNTs *J. Phys. Chem. C* **119** 9568–73

[47] Jang C, Adam S, Chen J-H, Williams E D, Das Sarma S and Fuhrer M S 2008 Tuning the Effective Fine Structure Constant in Graphene: Opposing Effects of Dielectric Screening on Short- and Long-Range Potential Scattering *Phys. Rev. Lett.* **101** 146805

[48] Riedl C, Coletti C and Starke U 2010 Structural and electronic properties of epitaxial graphene on SiC(0 0 0 1): a review of growth, characterization, transfer doping and hydrogen intercalation *J. Phys. D. Appl. Phys.* **43** 374009

[49] Llobet E 2013 Gas sensors using carbon nanomaterials: A review *Sensors Actuators B Chem.* **179** 32–45

[50] Tokarczyk M, Kowalski G, Możdżonek M, Borysiuk J, Stępniewski R, Strupiński W, Ciepielewski P and Baranowski J M 2013 Structural investigations of hydrogenated epitaxial graphene grown on 4H-SiC (0001) *Appl. Phys. Lett.* **103** 241915

[51] Kowalski G, Tokarczyk M, Dąbrowski P, Ciepielewski P, Możdżonek M, Strupiński W and Baranowski J M 2015 New X-ray insight into oxygen intercalation in epitaxial graphene grown on 4H-SiC(0001) *J. Appl. Phys.* **117** 105301

[52] Ostler M, Fromm F, Koch R J, Wehrfritz P, Speck F, Vita H, Böttcher S, Horn K and Seyller T 2014 Buffer layer free graphene on SiC(0001) via interface oxidation in water vapor *Carbon N. Y.* **70** 258–65

[53] Norimatsu W and Kusunoki M 2014 Epitaxial graphene on SiC{0001}: advances and perspectives *Phys. Chem. Chem. Phys.* **16** 3501

[54] Xu P, Qi D, Schoelz J K, Thompson J, Thibado P M, Wheeler V D, Nyakiti L O, Myers-Ward R L, Eddy C R, Gaskill D K, Neek-Amal M and Peeters F M 2014 Multilayer graphene, Moiré patterns, grain boundaries and defects identified by scanning tunneling microscopy on the m-plane, non-polar surface of SiC *Carbon N. Y.* **80** 75–81

[55] Lauffer P, Emtsev K V., Graupner R, Seyller T and Ley L 2008 Atomic and electronic structure of few-layer graphene on SiC(0001) studied with scanning tunneling microscopy and spectroscopy *Phys. Rev. B* **77** 155426

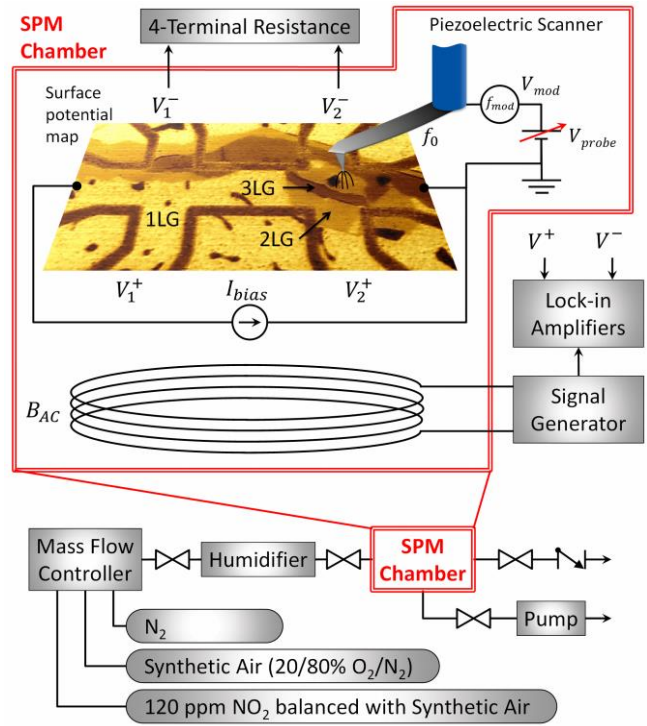

**Figure 1.** Schematics of the combined surface potential, transport measurements and environmental control system. Simultaneous surface potential and transport measurements are performed in the environmental SPM chamber, using FM-KPFM, AC Hall effect and channel resistance measurements. Central image is a surface potential map of the 1-3LG Hall bar device used in this experiment in vacuum with all contacts grounded.

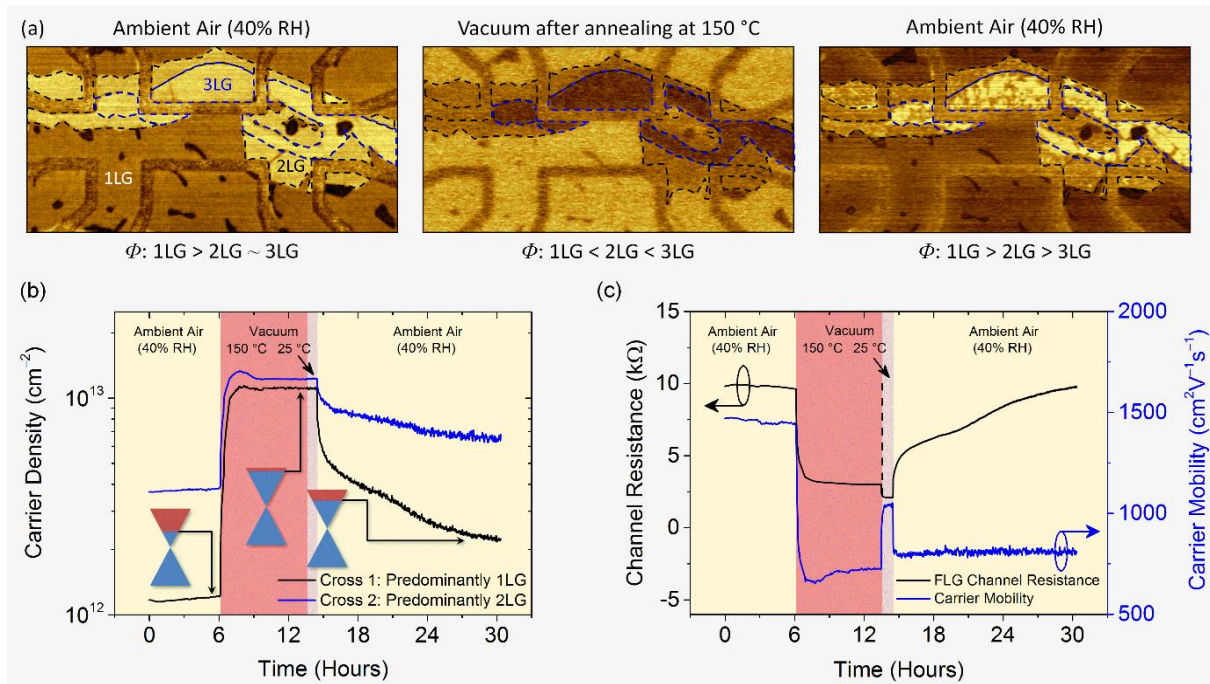

**Figure 2.** Surface potential and global transport properties of the Hall bar device in ambient air–vacuum annealing–ambient air. (a) Individual surface potential maps (400 mV colour scale) for ambient air, vacuum after annealing and ambient air. The scan size is 6×3 µm². Dashed lines mark 1-2-3LG domains. (b) 1LG and 2LG carrier densities, and (c) few-layer graphene (FLG) channel resistance and carrier mobility in the controlled environments. If not stated otherwise, all measurements were performed at

25 °C. Note: cross 1 is covered by 70/30% (1LG/2LG), cross 2 is 70/30% (2LG/3LG) and the FLG channel is 56/30/14% (1LG/2LG/3LG).

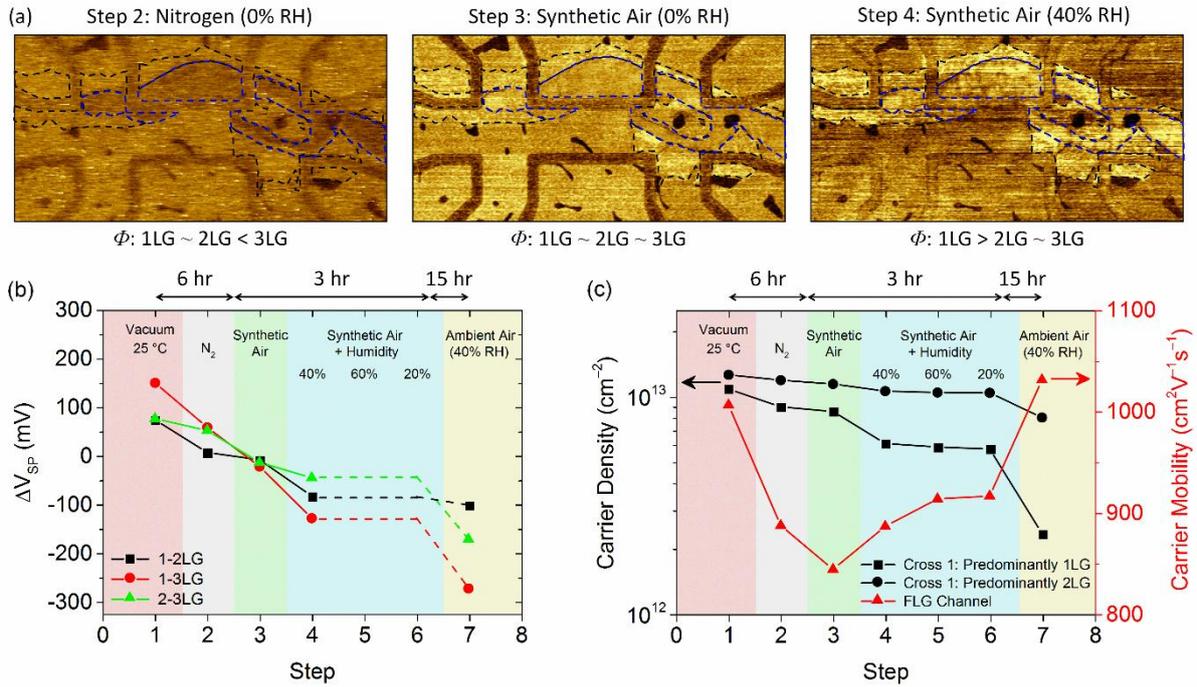

**Figure 3.** Measurements of surface potential and global transport in vacuum, $N_2$, synthetic air, 40-60-20% relative humidity and ambient air. (a) Individual surface potential maps (400 mV colour scale) for $N_2$, synthetic air and 40% relative humidity. The scan size is 6×3 μm$^2$. Dashed lines mark 1-2-3LG domains. (b) Surface potential difference ($\Delta V_{SP}$) between 1-2LG, 1-3LG and 2-3LG as derived from local surface potential maps. (c) Carrier density for 1LG and 2LG and carrier mobility for FLG channel in the controlled environments as derived from transport measurements. Note: the correlation between local ($\Delta V_{SP}$) and global ($n_e^{1LG}$ and $n_e^{2LG}$) properties.

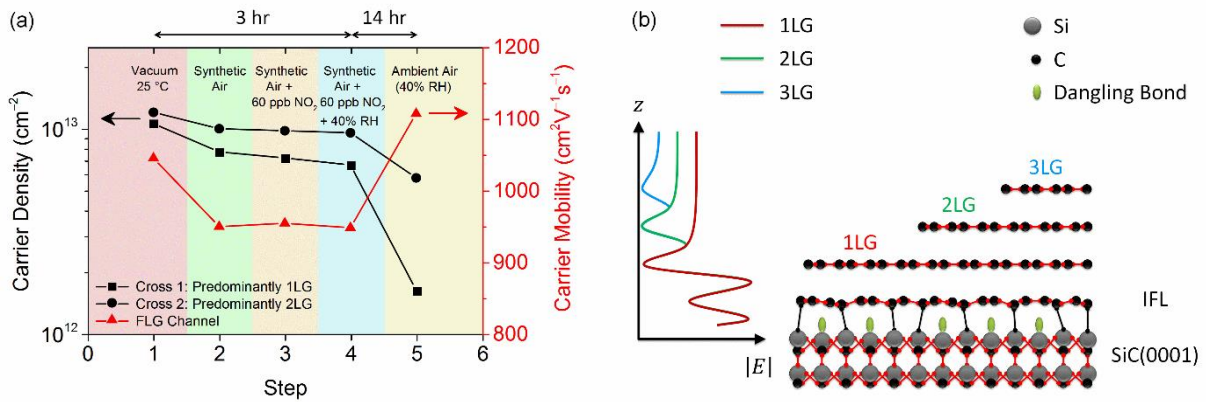

**Figure 4.** (a) Measurements of 1LG and 2LG carrier density and FLG carrier mobility in vacuum, synthetic air, 60 ppb $NO_2$, 60 ppb $NO_2$ with 40% relative humidity and ambient air. (b) Schematic representation of the electric field between the epitaxial graphene layers as a result of n-doping from the interfacial layer (IFL).

**Table 1.** Summary of the percentage change in carrier density and mobility relative to the pristine annealed vacuum state from Figure 3c, where SA is synthetic air (20% $O_2$ balanced with $N_2$).

| Environmental Condition | $\Delta n_e^{ILG}$ | $\Delta n_e^{2LG}$ | $\Delta n_e^{ILG}/\Delta n_e^{2LG}$ | $\Delta \mu_e$ |
|---|---|---|---|---|
| $N_2$ | −17.3% | −5.6% | 3.1 | −11.8% |
| SA | −21.2% | −9.4% | 2.3 | −16.2% |
| SA + 40% RH | −43.9% | −16.0% | 2.7 | −11.9% |
| SA + 60% RH | −46.0% | −17.2% | 2.6 | −9.2% |
| SA + 20% RH | −47.1% | −17.4% | 2.7 | −8.9% |
| Ambient Air | −78.4% | −36.5% | 2.1 | +2.5% |